\documentclass[conference]{IEEEtran}
\usepackage{tikz}
\usepackage{pgfplots}
\usepackage{booktabs}
\usepackage{amsmath}
\pgfplotsset{compat=1.17}
\begin{document}
\title{Res-MIA: A Training-Free Resolution-Based Membership Inference Attack on Federated Learning Models}

\author{
Mohammad~Zare \\ 
Department of Computer Engineering\\ and Information Technology \\
Shiraz University of Technology, Shiraz, Iran \\
\texttt{md.zare@sutech.ac.ir}
\and
Pirooz~Shamsinejadbabaki \\ 
Department of Computer Engineering\\ and Information Technology \\
Shiraz University of Technology, Shiraz, Iran \\
\texttt{p.shamsinejad@sutech.ac.ir}
}

\maketitle

\begin{abstract}
Membership inference attacks (MIAs) pose a serious threat to the privacy of machine learning models by allowing adversaries to determine whether a specific data sample was included in the training set. Although federated learning (FL) is widely regarded as a privacy-aware training paradigm due to its decentralized nature, recent evidence shows that the final global model can still leak sensitive membership information through black-box access. In this paper, we introduce Res-MIA, a novel training-free and black-box membership inference attack that exploits the sensitivity of deep models to high-frequency input details. Res-MIA progressively degrades the input resolution using controlled downsampling and restoration operations, and analyzes the resulting confidence decay in the model’s predictions. Our key insight is that training samples exhibit a significantly steeper confidence decline under resolution erosion compared to non-member samples, revealing a robust membership signal. Res-MIA requires no shadow models, no auxiliary data, and only a limited number of forward queries to the target model. We evaluate the proposed attack on a federated ResNet-18 trained on CIFAR-10, where it consistently outperforms existing training-free baselines and achieves an AUC of up to 0.88 with minimal computational overhead. These findings highlight frequency-sensitive overfitting as an important and previously underexplored source of privacy leakage in federated learning, and emphasize the need for privacy-aware model designs that reduce reliance on fine-grained, non-robust input features.
\end{abstract}

\begin{IEEEkeywords}
Membership Inference Attack, Federated Learning, Privacy Leakage, Black-Box Attacks, Model Overfitting, Image Resolution, Security in Machine Learning
\end{IEEEkeywords}

\section{Introduction}
Machine learning models can unintentionally reveal sensitive information about the data used during training, which raises serious privacy concerns. One of the most well-known threats in this context is the membership inference attack (MIA), where an adversary attempts to determine whether a particular data sample was included in the training set of a target model \cite{Shokri2017, Yeom2018}. This risk becomes more pronounced when models overfit, as they tend to memorize fine-grained and sample-specific characteristics rather than learning generalizable patterns \cite{Yeom2018, Feldman2020}. Although federated learning (FL) was introduced to reduce direct data sharing by enabling collaborative training over decentralized datasets \cite{McMahan2017, Kairouz2021}, it does not eliminate the threat of MIAs. An attacker with query access to the final global model can still exploit its outputs to infer training membership and compromise data privacy \cite{Nasr2019}. Prior work has shown that such attacks are effective not only in centralized learning settings \cite{Salem2019, Sablayrolles2019} but also in federated systems \cite{Nasr2019, Suri2023}, emphasizing the need for a deeper understanding of privacy leakage mechanisms.

Early membership inference methods relied on training shadow models to replicate the behavior of the target model and generate labeled data for an attack classifier \cite{Shokri2017}. While effective, these approaches are computationally expensive and depend on access to auxiliary data drawn from a similar distribution. Later studies proposed simpler alternatives that directly exploit the target model’s output statistics. For instance, loss-based attacks use prediction confidence or loss values as indicators of membership, under the assumption that training samples tend to produce lower loss or higher confidence \cite{Yeom2018}. Other methods rely on measures such as output entropy or confidence gaps between predicted classes \cite{Salem2019}. More recent label-only attacks operate under even more restricted conditions, assuming access only to the predicted label rather than full confidence scores, and infer membership by probing the decision boundary through input variations \cite{Chen2021, Wu2024}. Although these approaches reduce information requirements, they often require many queries or carefully crafted inputs. In federated learning, active attacks have also been explored, where a malicious participant or server manipulates the training process to amplify membership signals \cite{Nasr2019, Nguyen2023}. Additionally, some studies have investigated membership inference at coarser levels, such as subject-level inference across federated clients \cite{Suri2023}. Despite this progress, the potential of using systematic input transformations to expose membership leakage in a purely black-box and training-free manner has remained largely unexplored.

In this work, we introduce Res-MIA, a black-box membership inference attack that does not require training auxiliary attack models or shadow networks. The core intuition behind our approach is that models tend to rely on high-frequency and fine-grained details when making predictions on training samples, while their predictions on unseen samples are more strongly driven by coarse, low-frequency features. By progressively removing high-frequency content from the input through controlled resolution degradation, we observe that the model’s confidence on member samples decreases much more rapidly than on non-member samples. Based on this observation, we propose a simple and efficient attack pipeline that measures the rate of confidence decay across successive input transformations. Our method operates entirely in a black-box setting, requires only a small number of forward passes, and introduces a confidence decay score to quantify membership likelihood. Experimental results on a federated ResNet-18 model trained on CIFAR-10 show that Res-MIA achieves substantially higher attack performance than existing training-free baselines, while remaining computationally lightweight. We also analyze key design choices in the attack pipeline and show that preserving coarse artifacts during input degradation plays an important role in amplifying membership signals.

The remainder of this paper is organized as follows. Section II reviews related work on membership inference attacks and privacy risks in federated learning. Section III describes the proposed Res-MIA method in detail, including the progressive input erosion process and the confidence decay metric. Section IV presents experimental results and ablation studies that evaluate the effectiveness of the attack. Finally, Section V concludes the paper and discusses implications for privacy-preserving model design and future research directions.

\section{Related Work}
Membership inference attacks were first introduced by Shokri et al. \cite{Shokri2017}, who proposed training multiple shadow models to imitate the behavior of a target model. The outputs of these shadow models are then used to train a separate attack classifier that distinguishes between member and non-member samples. Although effective, this line of work requires training several auxiliary models and assumes access to data drawn from a distribution similar to that of the target model. Later studies proposed refinements to reduce the amount of required shadow data or improve attack robustness \cite{Salem2019, Truex2019}, but the overall computational overhead of shadow-model-based approaches remains high.

To address these limitations, subsequent research explored simpler attacks that do not rely on training auxiliary models. Yeom et al. \cite{Yeom2018} demonstrated that overfitting alone can be exploited by directly thresholding the model’s prediction confidence or loss on a given input. Samples that produce very high confidence or low loss are more likely to belong to the training set. Related work showed that entropy of the output probability distribution can also serve as a membership signal, where lower entropy indicates a higher likelihood of membership \cite{Salem2019}. These one-shot attacks are computationally inexpensive and easy to deploy, but their effectiveness is often limited because they rely on a single model response and fail to capture richer behavioral differences between member and non-member samples.

More recent studies consider scenarios in which the adversary has restricted access to the model outputs. In label-only attacks, the attacker observes only the predicted class label without access to confidence scores. To compensate for this limitation, these methods probe the model’s decision boundary by applying perturbations or adversarial noise to the input and observing changes in the predicted label \cite{Chen2021}. Wu et al. \cite{Wu2024} proposed the You Only Query Once (YOQO) attack, which uses a carefully crafted augmented query to infer membership with significantly fewer queries than earlier label-only approaches. While label-only attacks reduce the information requirements on the attacker, they often depend on the ability to generate meaningful input variations, and their performance can vary depending on how easily such variations can be constructed.

Federated learning introduces additional perspectives on membership inference. Nasr et al. \cite{Nasr2019} showed that both passive and active attacks are possible in federated settings, demonstrating that membership information can be inferred even when only model updates are exchanged. Follow-up work explored user-level membership inference, where the goal is to determine whether a particular client participated in training, as well as defenses such as secure aggregation that aim to hide individual updates. More recently, Suri et al. \cite{Suri2023} studied subject-level membership inference in cross-silo federated learning, where data from a single individual may be distributed across multiple organizations. Their results showed that aggregating signals over training rounds can enable effective inference even under certain privacy defenses. In contrast, our work focuses on the standard item-level membership setting and assumes access only to the final global model.

A parallel body of work has focused on defending against membership inference attacks. Differential privacy provides formal guarantees against information leakage by injecting noise during training \cite{Dwork2006}, but often at the cost of reduced model accuracy when strong privacy budgets are enforced. Other approaches incorporate adversarial regularization or training procedures that explicitly reduce the distinguishability between member and non-member samples \cite{Nasr2018}. Post-processing defenses such as MemGuard add carefully designed noise to model outputs at inference time to mislead attackers \cite{Jia2019}. While these defenses can be effective in certain settings, they typically require careful tuning and may negatively impact utility. By highlighting the role of high-frequency overfitting in membership leakage, our work complements this literature and suggests that regularizing models to reduce reliance on fine-grained input details may serve as an additional defense mechanism.

Table~\ref{tab:comparison} provides a summary of key differences between representative membership inference attack paradigms and the proposed Res-MIA method.

\begin{table*}[!ht]
\caption{Comparison of membership inference attack methods.\label{tab:comparison}}
\centering
\begin{tabular}{lccc}
\toprule
Attack Method & Comp. Cost & Assumptions & Access Level \\
\midrule
Shadow-model Attack \cite{Shokri2017, Salem2019} & High & Shadow data and model training & Black-box (confidences) \\
Entropy/Loss Attack \cite{Yeom2018, Salem2019} & Low & No training; one query & Black-box (confidences) \\
Label-only Attack \cite{Chen2021, Wu2024} & Low & No confidences; multi-query & Black-box (label-only) \\
Res-MIA (Ours) & Low & No training; iterative input modification & Black-box (confidences) \\
\bottomrule
\end{tabular}
\vspace{-1em}
\end{table*}

\section{Methodology}
In this section, we describe the proposed attack methodology and formalize the intuition behind it. The goal is to characterize how a trained model responds to systematic input degradation and to quantify this response in a way that reveals membership information. To this end, we first introduce the progressive erosion process and then define a confidence-based metric that captures the sensitivity of model predictions to resolution loss.

\subsection{Progressive Image Erosion}
The core idea of the proposed attack is to progressively suppress high-frequency information in the input and observe how the model’s prediction behavior changes as a result. This process is motivated by recent findings showing that overfitted models tend to rely heavily on fine-grained details, which are more prevalent in training samples than in unseen data \cite{Pang2022, Zhang2023Frequency}. By gradually removing these details, we aim to expose differences in prediction stability between member and non-member samples.

Starting from an original input image $x_0$, we apply a sequence of $K$ erosion steps. Each step consists of reducing the spatial resolution of the image and then restoring it back to the original size. Average pooling is used for downsampling, followed by nearest-neighbor interpolation for upsampling. Let $f(\cdot)$ denote the target classification model, which outputs a probability distribution over classes. The erosion process is defined as
\begin{equation}
x_k = U\big(\mathrm{AvgPool}(x_{k-1})\big), \qquad k = 1,2,\ldots,K ,
\end{equation}
where $x_0$ is the original input, $\mathrm{AvgPool}(\cdot)$ reduces the image resolution, and $U(\cdot)$ restores the image to its original size. After several iterations, the eroded image $x_K$ preserves only coarse spatial structures and color information, while most high-frequency details are removed. Similar forms of input degradation have been shown to reveal robustness and privacy-related properties of deep models \cite{He2024Robustness}.

\begin{figure*}[!ht]
\centering
\includegraphics[width=0.95\textwidth]{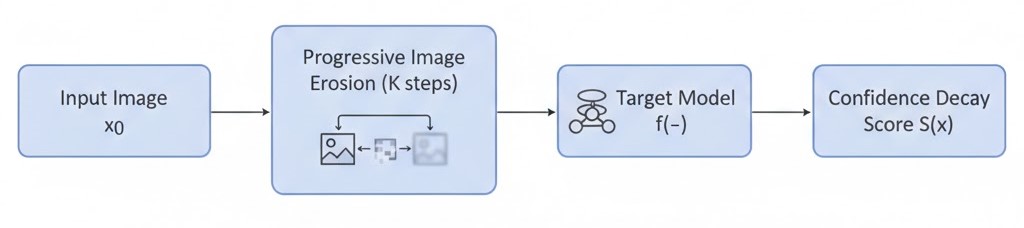}
\caption{Overview of the Res-MIA attack pipeline. The input image is progressively degraded through resolution erosion. The target model is queried at each erosion step, and the resulting confidence decay is used to infer membership.}
\label{fig:pipeline}
\end{figure*}

Figure~\ref{fig:pipeline} illustrates a high-level overview of the attack pipeline. The input image is progressively eroded, queried by the target model at each stage, and summarized using a confidence-based score that reflects prediction sensitivity.

At each erosion level $k$, the model is queried to obtain its prediction confidence. Let
\begin{equation}
c_k = \max_y f_y(x_k)
\end{equation}
denote the confidence associated with the predicted class at erosion step $k$. While training samples often produce higher initial confidence values due to memorization effects \cite{Carlini2023, Liu2022FLPrivacy}, the primary signal exploited by our method lies in the evolution of the confidence sequence $\{c_0, c_1, \ldots, c_K\}$. For member samples, confidence typically drops rapidly as fine-grained details are removed. In contrast, non-member samples tend to exhibit a more gradual decline, as their predictions rely more on robust and general features.

To quantify this behavior, we define the confidence decay score, which summarizes the rate at which the model’s confidence decreases over erosion steps. Let $y^* = \arg\max_y f_y(x_0)$ be the predicted class for the original input. The confidence decay score is defined as
\begin{equation}
S(x) = \frac{1}{K} \sum_{k=1}^{K} \left( f_{y^*}(x_{k-1}) - f_{y^*}(x_k) \right).
\end{equation}
This score represents the average confidence loss per erosion step. It can also be expressed as
\begin{equation}
S(x) = \frac{f_{y^*}(x_0) - f_{y^*}(x_K)}{K},
\end{equation}
which highlights that the score captures the total confidence decay normalized by the number of erosion steps. Larger values of $S(x)$ indicate higher sensitivity to input degradation and, consequently, a higher likelihood that the sample belongs to the training set. This form of sensitivity-based analysis is closely related to recent studies on frequency bias and spectral properties of deep networks \cite{Zhang2023Frequency, Zhao2024Spectral}.

\begin{figure}[!ht]
\centering
\includegraphics[width=0.9\columnwidth]{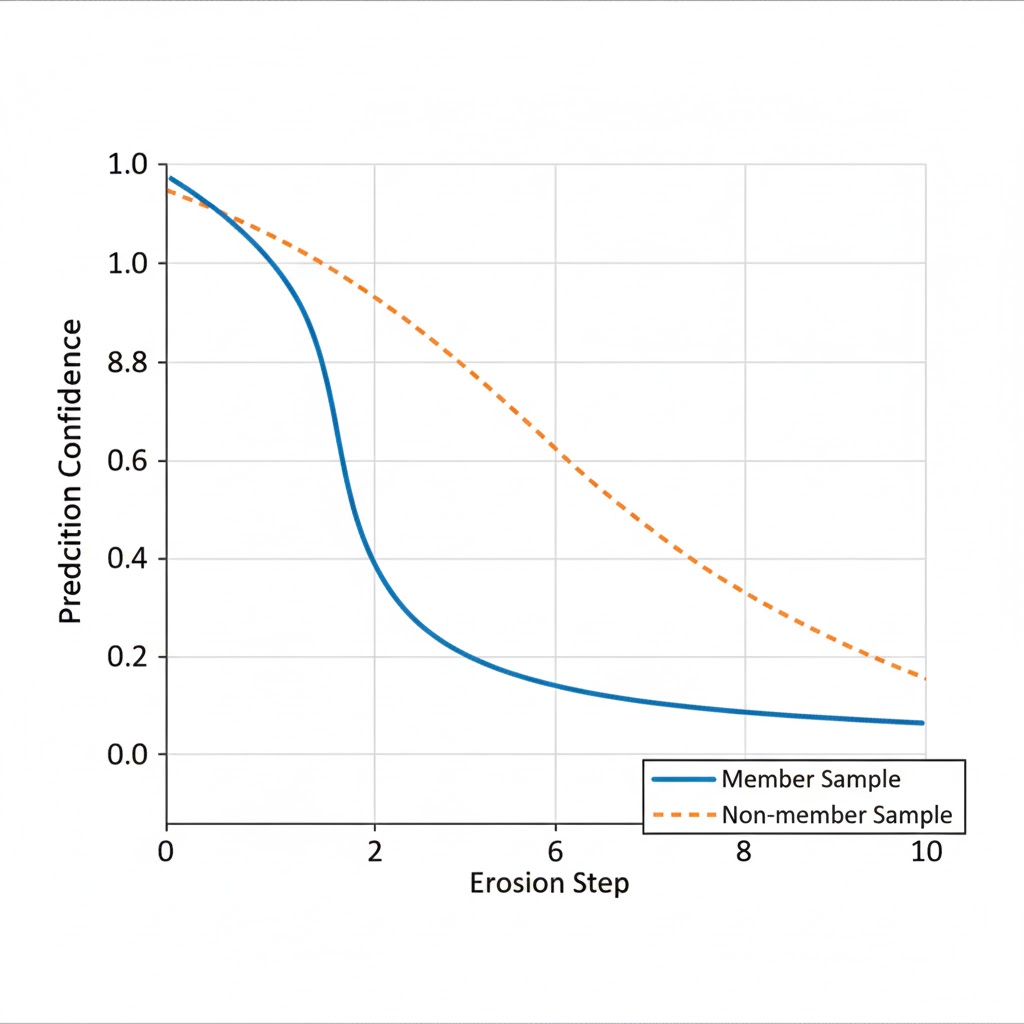}
\caption{Illustrative confidence decay curves under progressive image erosion. Member samples exhibit a steeper confidence decline compared to non-member samples.}
\label{fig:decay}
\end{figure}

Figure~\ref{fig:decay} provides an illustrative example of confidence decay behavior for member and non-member samples under progressive erosion.

\subsection{Design Rationale for Pooling and Upsampling}
The choice of average pooling and nearest-neighbor upsampling is motivated by their complementary effects on input structure. Average pooling acts as a low-pass filter, suppressing high-frequency content while preserving coarse spatial patterns. Nearest-neighbor upsampling restores the image size without introducing new pixel values, ensuring that removed details are not artificially reconstructed. In contrast, smoother interpolation methods such as bilinear or bicubic interpolation tend to reintroduce intermediate values that partially mask the effects of degradation.

This combination creates a controlled and monotonic degradation path for the input. Models that rely on brittle, high-frequency cues become increasingly uncertain as erosion progresses, while models that base their predictions on more stable features retain confidence for a longer range of degradation. Importantly, the proposed method requires no retraining, gradient access, or modification of the target model, and relies solely on forward passes. These properties make the attack simple to deploy and applicable in strict black-box settings, consistent with recent analyses of training-free membership inference \cite{Carlini2023}.

\subsection{Experimental Setup}
We evaluate the proposed Res-MIA attack on a standard image classification task in a federated learning setting using the CIFAR-10 dataset. CIFAR-10 contains 60,000 color images of size $32\times32$ spanning 10 object classes, with 50,000 samples used for training and 10,000 for testing. Due to its moderate scale and diverse visual content, CIFAR-10 is widely adopted for analyzing both generalization behavior and privacy leakage in deep learning models \cite{Kairouz2021, Choquette2023FL}.

\begin{figure*}[!ht]
\centering
\includegraphics[width=\linewidth]{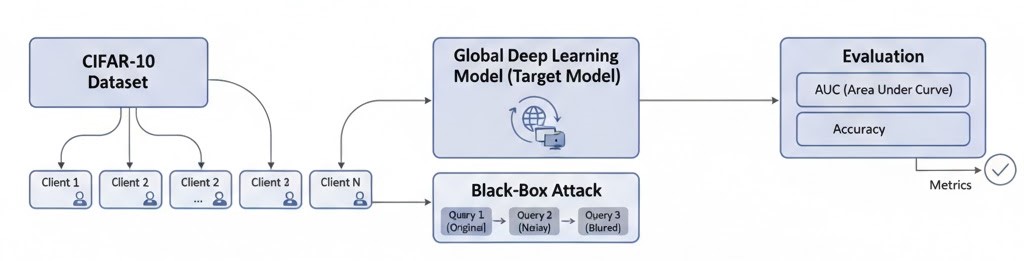}
\caption{Experimental evaluation pipeline of Res-MIA in a federated learning setting.}
\label{fig:exp_overview}
\end{figure*}

To emulate a federated learning environment, the training set is evenly partitioned across $N=10$ clients, resulting in 5,000 local training samples per client. A ResNet-18 architecture \cite{He2016} is trained using the FedAvg algorithm \cite{McMahan2017} for a fixed number of communication rounds. After convergence, the resulting global model achieves a test accuracy of approximately 85\%, which aligns with prior empirical studies under comparable federated configurations. The trained global model is treated as a black-box and constitutes the target of the membership inference attack. An overview of the experimental pipeline is illustrated in Figure~\ref{fig:exp_overview}.

For attack evaluation, we construct a balanced membership inference dataset consisting of 2,000 samples. The member set contains 1,000 images randomly selected from the federated training data, with equal contributions from all clients to mitigate client-specific bias. The non-member set consists of 1,000 images drawn from the CIFAR-10 test split, which are guaranteed not to have been observed during training. The attacker’s objective is to distinguish between these two groups solely based on the target model’s outputs, consistent with realistic black-box threat models considered in prior work \cite{Nasr2021Practical, Zhang2024Benchmark}.

\begin{figure}[!ht]
\centering
\includegraphics[width=\linewidth]{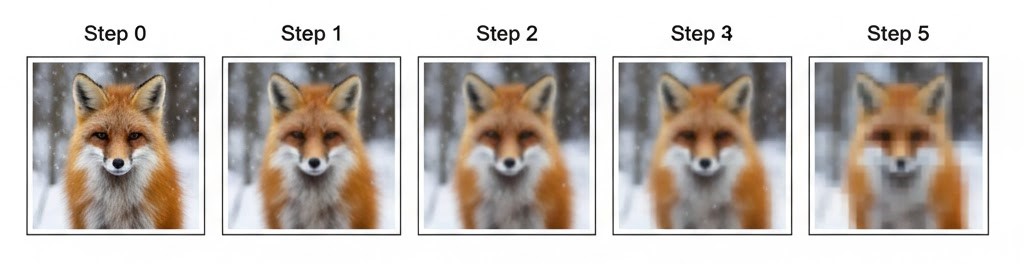}
\caption{Examples of progressive image erosion through successive resolution degradation steps.}
\label{fig:eroded_examples}
\end{figure}

Unless stated otherwise, we use $K=5$ erosion steps in Res-MIA. Starting from the original resolution of $32\times32$, successive $2\times$ downsampling operations reduce the image to an effective resolution of $1\times1$, followed by upsampling back to the original size. This aggressive degradation setting was empirically chosen to remove nearly all high-frequency details while preserving coarse color statistics. Examples of progressively eroded inputs are shown in Figure~\ref{fig:eroded_examples}.

\subsection{Evaluation Metrics}
Attack performance is primarily evaluated using the area under the receiver operating characteristic curve (AUC), a threshold-independent metric that captures the trade-off between true positive rate and false positive rate. AUC is particularly well-suited for membership inference attacks, where the operating threshold may vary across deployments \cite{Song2022Confidence}. In addition to AUC, we report classification accuracy, the false positive rate at a fixed true positive rate of 80\%, and computational overhead to provide a comprehensive assessment of attack effectiveness.

\subsection{Attack Performance and Comparative Analysis}
\begin{figure}[!ht]
\centering
\includegraphics[width=0.9\linewidth]{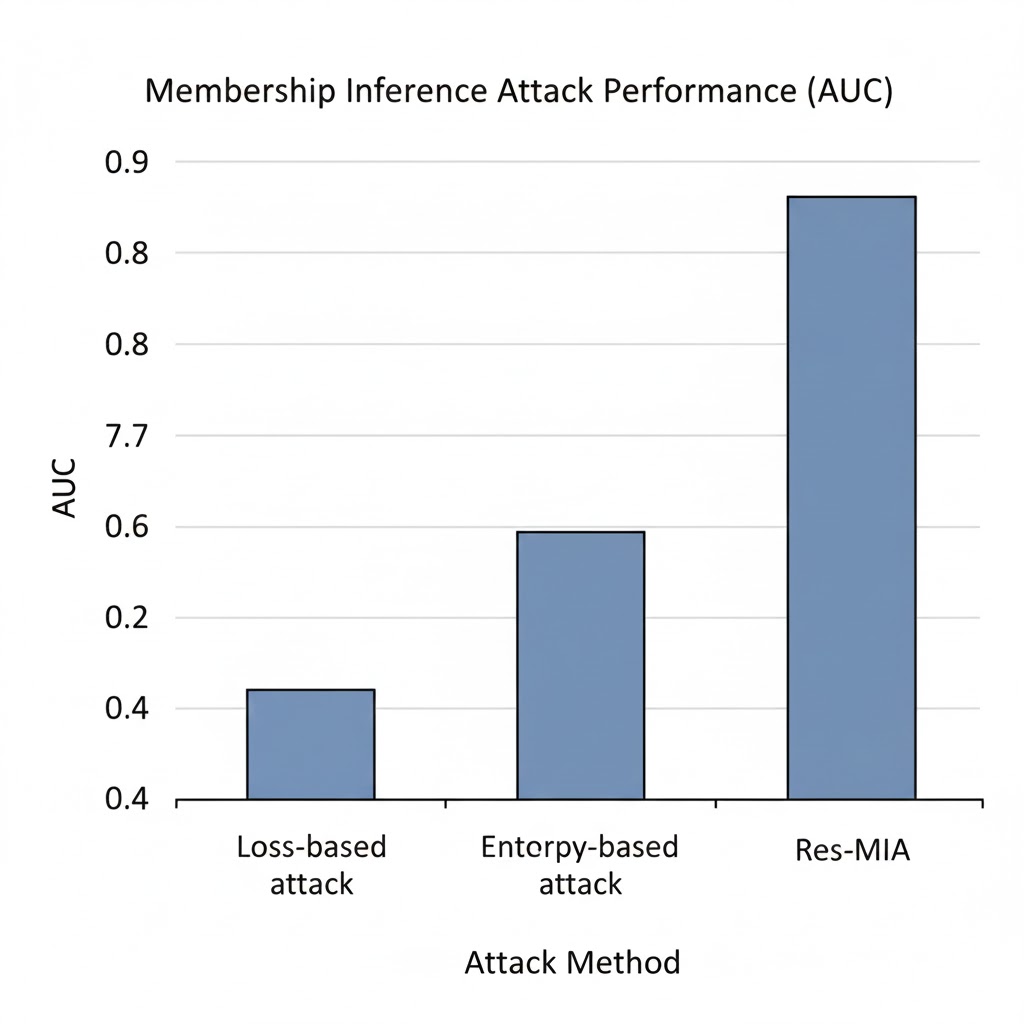}
\caption{Comparison of AUC values for different membership inference attacks.}
\label{fig:auc_comparison}
\end{figure}

We compare Res-MIA against two widely adopted training-free baseline attacks: a loss-based attack that thresholds prediction confidence \cite{Yeom2018} and an entropy-based attack that relies on output distribution entropy \cite{Salem2019}. Quantitative results are summarized in Table~\ref{tab:performance_main}, while a visual comparison of AUC values is provided in Figure~\ref{fig:auc_comparison}.

\begin{table}[!ht]
\caption{Membership inference performance on federated CIFAR-10 using a ResNet-18 model.\label{tab:performance_main}}
\centering
\begin{tabular}{lccc}
\toprule
Attack Method & AUC & Accuracy & FPR @ TPR=80\% \\
\midrule
Loss-Based Attack \cite{Yeom2018} & 0.75 & 0.69 & 0.38 \\
Entropy-Based Attack \cite{Salem2019} & 0.68 & 0.64 & 0.45 \\
Res-MIA (Ours) & 0.88 & 0.81 & 0.19 \\
\bottomrule
\end{tabular}
\end{table}

Res-MIA consistently outperforms both baselines across all reported metrics. Notably, at a fixed true positive rate of 80\%, the false positive rate is reduced by nearly half compared to confidence-based attacks. This reduction is particularly important in practical privacy auditing scenarios, where falsely identifying non-member samples as members can have serious implications \cite{Nasr2021Practical}. The results confirm that confidence decay under progressive input erosion provides a substantially stronger membership signal than single-query confidence or entropy measures.

\subsection{Client-Wise Robustness Analysis}
\begin{figure}[!ht]
\centering
\includegraphics[width=0.9\linewidth]{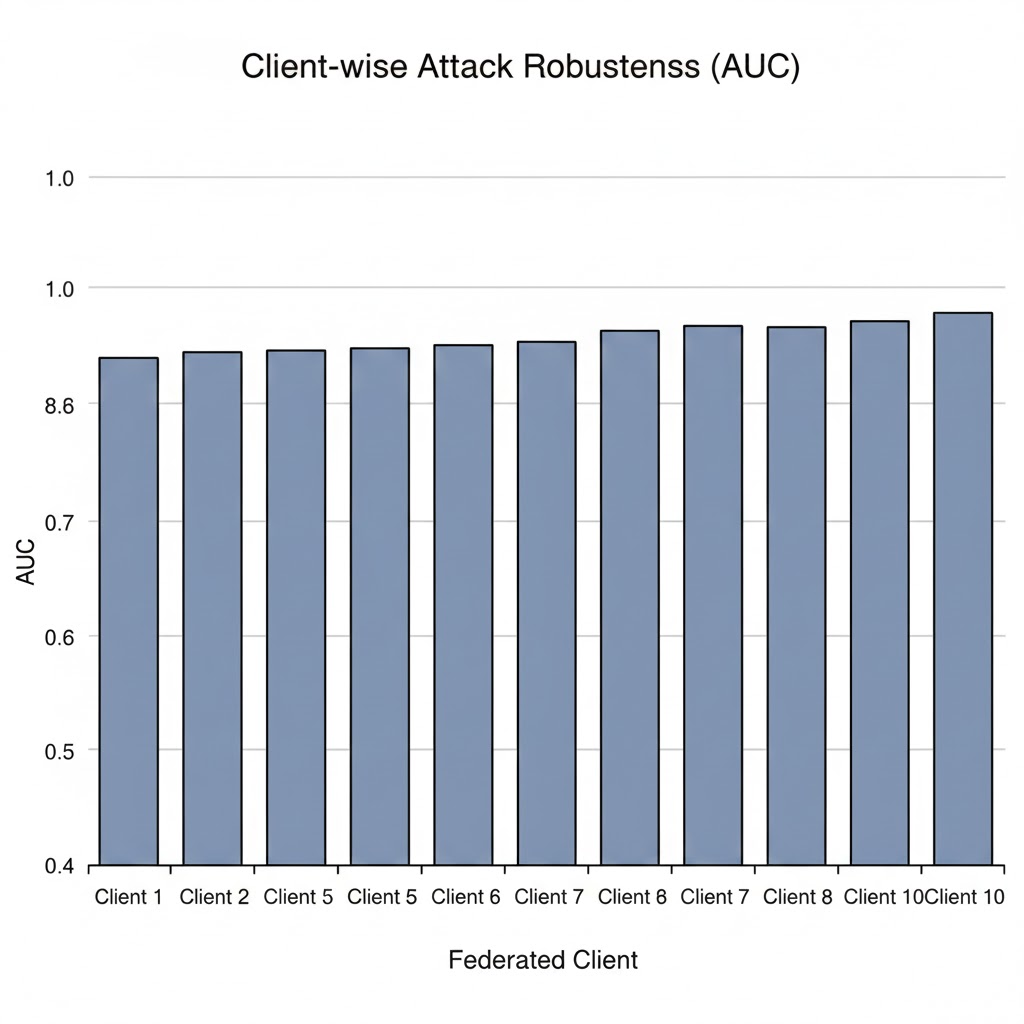}
\caption{Client-wise AUC performance of Res-MIA across federated participants.}
\label{fig:clientwise_auc}
\end{figure}

To examine whether the attack performance is consistent across different federated clients, we evaluate Res-MIA separately on member samples originating from each client. The resulting client-wise AUC values are reported in Table~\ref{tab:clientwise} and visualized in Figure~\ref{fig:clientwise_auc}.

\begin{table}[!ht]
\caption{Client-wise AUC of Res-MIA across federated participants.\label{tab:clientwise}}
\centering
\begin{tabular}{cccccc}
\toprule
Client 1 & Client 2 & Client 3 & Client 4 & Client 5 \\
\midrule
0.87 & 0.88 & 0.86 & 0.89 & 0.88 \\
\midrule
Client 6 & Client 7 & Client 8 & Client 9 & Client 10 \\
\midrule
0.87 & 0.88 & 0.89 & 0.86 & 0.88 \\
\bottomrule
\end{tabular}
\end{table}

The relatively small variance across clients indicates that Res-MIA does not exploit client-specific artifacts and remains effective under uniform data partitioning, which is consistent with recent benchmarking studies on federated membership inference \cite{Zhang2024Benchmark}.

\subsection{Ablation Study on Upsampling Strategy}
We further investigate the impact of the upsampling method used during erosion. Specifically, we compare nearest-neighbor upsampling with bilinear interpolation while keeping all other components fixed. The results are reported in Table~\ref{tab:ablation_up} and illustrated in Figure~\ref{fig:ablation_plot}.
\begin{figure}[!ht]
\centering
\includegraphics[width=0.9\linewidth]{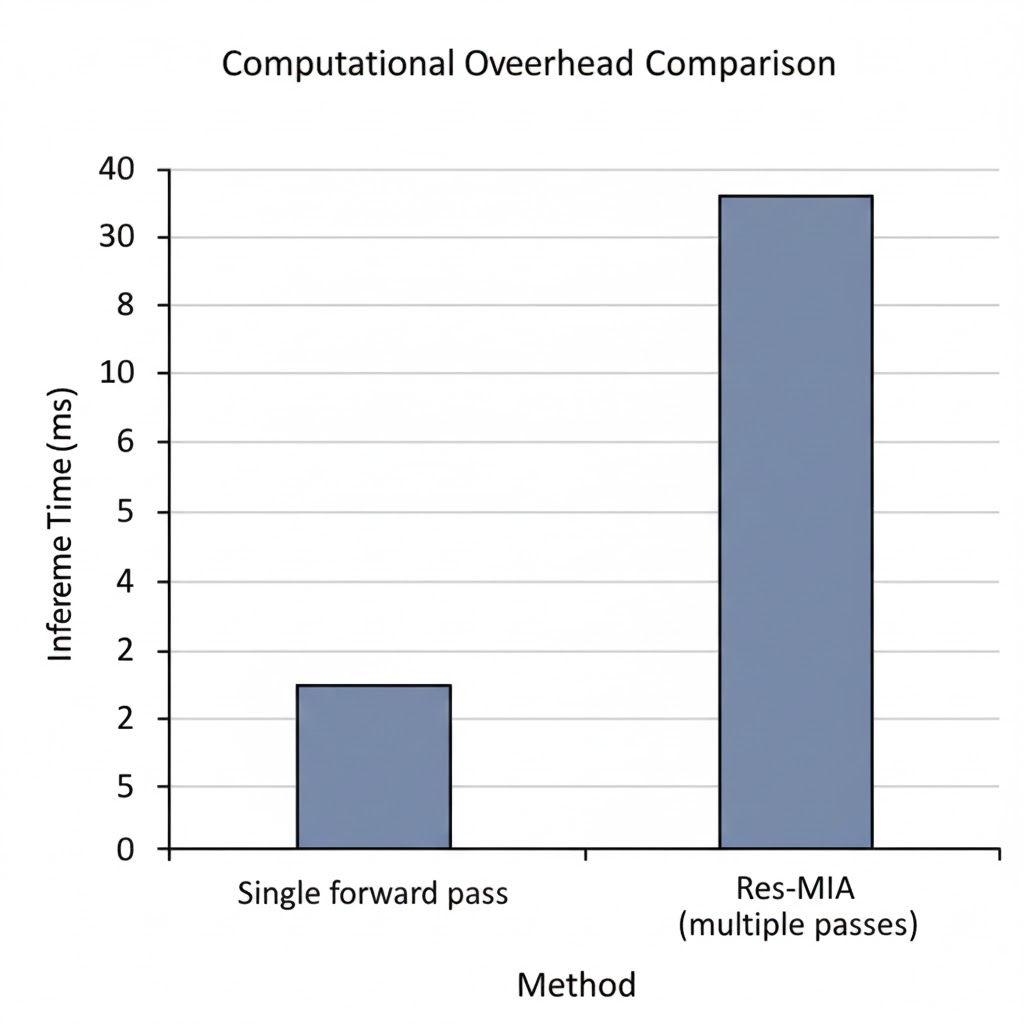}
\caption{Computational overhead comparison between a single forward pass and the Res-MIA attack.}
\label{fig:overhead_plot}
\end{figure}

\begin{table}[!t]
\caption{Effect of upsampling method on attack performance.\label{tab:ablation_up}}
\centering
\begin{tabular}{lc}
\toprule
Upsampling Method & AUC \\
\midrule
Nearest-Neighbor & 0.88 \\
Bilinear Interpolation & 0.72 \\
\bottomrule
\end{tabular}
\end{table}

The substantial performance degradation observed with bilinear interpolation suggests that smoother reconstruction partially obscures the effects of high-frequency removal. In contrast, nearest-neighbor upsampling preserves coarse, block-like artifacts that amplify confidence decay differences between member and non-member samples.

\subsection{Computational Overhead}
Finally, we analyze the computational overhead of Res-MIA. For each queried sample, the attacker performs $K+1$ forward passes through the target model. With $K=5$, this results in six inferences per sample. Average inference times are reported in Table~\ref{tab:overhead} and summarized visually in Figure~\ref{fig:overhead_plot}.

\begin{table}[!t]
\caption{Computational overhead of Res-MIA per queried sample.\label{tab:overhead}}
\centering
\begin{tabular}{lc}
\toprule
Method & Inference Time (ms) \\
\midrule
Single Forward Pass & 3.1 \\
Res-MIA ($K=5$) & 18.9 \\
\bottomrule
\end{tabular}
\end{table}

Despite the additional queries, the overhead remains modest and fully parallelizable, making Res-MIA practical for large-scale or online attack scenarios.

\section{Conclusion}
In this work, we introduced Res-MIA, a training-free membership inference attack that leverages the sensitivity of deep models to high-frequency input details. By progressively degrading the resolution of input images and tracking the resulting confidence decay, Res-MIA is able to reliably distinguish training samples from unseen data using only black-box access to the final model. Experimental results in a federated learning setting demonstrate that confidence decay provides a strong and consistent membership signal, outperforming commonly used one-shot baselines while remaining computationally lightweight.

Beyond its empirical effectiveness, Res-MIA offers insights into the underlying causes of privacy leakage in modern deep learning models. Our findings suggest that memorization of fine-grained, high-frequency details plays a critical role in exposing membership information, particularly in over-parameterized models trained in distributed settings. This observation indicates that mitigating membership inference risks requires more than improving generalization performance; it calls for explicitly reducing a model’s reliance on brittle, detail-specific cues that are not essential for robust prediction.

Looking ahead, several directions emerge from this work. One promising avenue is the development of training strategies that explicitly suppress sensitivity to high-frequency perturbations, thereby weakening the confidence decay patterns exploited by Res-MIA. Another direction is to extend the core idea of progressive erosion to other data modalities, such as text or higher-resolution visual inputs, to study whether similar membership signals arise. Finally, combining confidence decay with complementary attack signals could enable more powerful inference at finer granularity, such as user-level membership in federated systems. We believe these directions will contribute to a deeper understanding of the relationship between model memorization, robustness, and privacy.

\bibliographystyle{IEEEtran}
\bibliography{references}

\end{document}